# Theory of water desalination by porous electrodes with fixed chemical charge

P.M. Biesheuvel,[1,2] M.E. Suss,[3,*] and H.V.M. Hamelers[1]

[1]*Wetsus, European Centre of Excellence for Sustainable Water Technology, Leeuwarden, The Netherlands.*

[2]*Laboratory of Physical Chemistry and Soft Matter, Wageningen University, The Netherlands.* [3]*Faculty of Mechanical Engineering, Technion - Israel Institute of Technology, Haifa, Israel.*

*Corresponding author. E-mail: mesuss@technion.ac.il

**Abstract**

Water desalination by capacitive deionization (CDI) is performed via electrochemical cells consisting of two porous carbon electrodes. Upon transferring charge from one electrode to the other, ions are removed from the feedwater by electrosorption into electrical double layers (EDLs) within the micropores of the porous carbon. When using electrodes containing fixed chemical charge in the micropores, various counterintuitive observations have been made, such as "inverted CDI" where upon charging, ions are released from the electrode, and the feedwater is only desalinated when the cell is discharging. We set up an EDL model including chemical charge that explains these observations and makes predictions for a working range of enhanced desalination by CDI.

------

In Capacitive Deionization (CDI), water is desalinated by the phenomenon of electrosorption in porous carbon electrodes.[1-4] Upon transferring electronic charge through an external circuit from one porous electrode to the other, electrical double layers (EDLs) are formed in the micropores within the carbon electrodes. In micropores, electronic charge (in the carbon) is locally charge-compensated by ionic charge, due to a difference in concentration between micropore counterions and coions, see Fig. 1A,B.[5-9] Because more counterions are adsorbed in the EDLs than coions are expelled, the water flowing through the CDI device is desalinated for a certain duration, typically of the order of minutes. When the electrodes become fully charged, the cell is discharged and ions are again released, leading to a brine stream. The most widely-used CDI cell architecture is called "flow-between" CDI where the two porous electrodes sandwich a spacer channel through which the feed water flows, but many other architectures have recently been demonstrated.[2] The measurable voltage between the two electrodes is the cell voltage $V_{cell}$ (anode minus cathode potential), and during CDI cell operation, the cell voltage switches between the charging voltage (for instance, $V_{ch}$=1.2 V), and the discharge voltage (typically $V_{disch}$=0). In this cyclical operational scheme, freshwater is produced for some time (during the charging step), after which for some time water of a higher salinity is produced (discharge step).

It is now well documented that it is possible for this classical response of a CDI cell to applied step voltages, to be inverted (i-CDI).[10-13] In i-CDI, during the charging step salt *desorbs* from the electrodes and the cell effluent is a brine. Ion electrosorption (desalination) occurs only in the cell's discharge step (when $V_{disch}$=0). This behavior is highly counterintuitive, and has been observed for electrodes modified to contain surface groups with a special chemical functionality,[11,13,14] or for those that during operation slowly developed oxygenated surfaces.[10,11,12] Various explanations for inverted operation have been proposed such as a shift in point of zero charge (pH$_{PZC}$),[13] potential of zero charge ($E_{PZC}$),[10,11,13] electrochemical oxidation of electrode and/or water,[10,11,13] pH effects,[11] coion desorption,[10,11] asymmetric potential distributions,[10,11] and chemical charge.[13] A quantitative theory which captures and explains this inversion phenomenon has not yet been proposed. Understanding the cause and implications of inversion is of importance for CDI and other processes using porous electrodes, and leveraging these effects can lead to improvements in CDI cell performance.

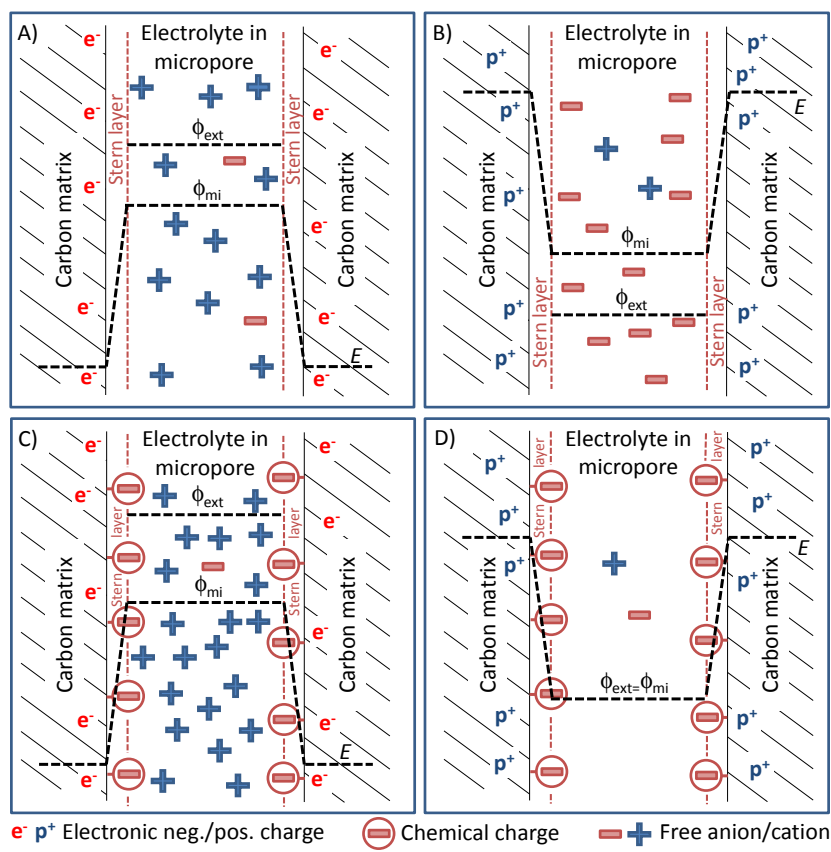

Fig. 1. Sketch of EDL structure in CDI for standard and modified electrodes. A) Standard EDL in electrode (cathode) where an excess of cations in the electrolyte phase in the micropore is charge-compensated by electronic charge in the carbon matrix. B) Reversing the charge has no effect on the total ion concentration. C) With fixed chemical charge residing in the cathode, more cations are adsorbed for the same electronic charge as in A). D). Reversing polarity leads to the expulsion of free cations, and the chemical charge being compensated by positive electronic charge. For equilibrium, the electrode potential $E$ is defined relative to the solution potential outside the micropore, $\phi_{ext}$. The Donnan potential is $\Delta\phi_D = \phi_{mi} - \phi_{ext}$; the Stern potential $\Delta\phi_S = E/V_T - \phi_{mi}$.

We here extend classical theory for porous electrodes by including a fixed chemical charge, which does not vary with liquid or solid phase potential, in the description of the EDL that forms in the micropores of the electrodes. Our EDL model considers electronic charge, mobile ionic charge, and fixed chemical charge, see Fig. 1C and D, taking inspiration from models capturing the semiconductor/water interface.[15-19] As we show, this extension of classical EDL theory captures the essential features observed in the performance of CDI with chemically modified electrodes, including accurately reproducing inverted operation. This agreement is compelling evidence that i-CDI is indeed the consequence of the presence of fixed chemical charge in the electrodes. In addition to describing i-CDI, our theory demonstrates the existence of new, unexplored operational modes. One resulting discovery we present here, is that for chemically modified electrodes operated in carefully chosen voltage ranges, the salt adsorption capacity in a CDI cycle (SAC) can be increased significantly compared to what has been previously achieved with unmodified capacitive electrodes.

The theory is based on a simple geometry of two porous carbon electrodes with internal, wetted micropore volume $v_{mi}$ per electrode. The electrodes are separated by an electrolyte-filled spacer channel. The electrolyte is a monovalent salt solution, such as NaCl in water. For classical CDI, the ionic charge in the EDLs, $\sigma_{ionic}$, exactly compensates the charge in the electronically conducting phase, i.e., in the carbon matrix, $\sigma_{elec}$ (Fig. 1A,B). For chemically modified electrodes, we include chemical charge in the charge balance, and thus $\sigma_{ionic}+\sigma_{chem}+\sigma_{elec}=0$ (Fig. 1C,D). Fixed chemical charge in porous carbons can be due to sulfonic groups (negative charge) or amine groups (positive charge).[20,21]

To describe the EDL structure we use the Donnan model, which is a simple model that captures the essential physics of salt and charge storage in capacitive porous electrodes.[5,6] Modified versions of the Donnan model accurately describe data for charge and salt adsorption in carbon micropores.[2,7,22] The Donnan model is based on the fact that in carbon micropores (< 2 nm), the EDLs forming along pore surfaces are strongly overlapped. Thus, a single electrostatic potential and concentration in the micropore electrolyte volume can be assumed, with the micropore ion concentration related to concentrations outside the micropore by the equilibrium Boltzmann relation, similar to its use for ion-exchange membranes.[23] The potential difference between inside and outside the micropore is the Donnan potential, $\Delta\phi_D$. An additional potential drop, $\Delta\phi_S$, is due to a constant capacitance, $C_S$, between electrolyte volume and carbon matrix. This capacitance may have its origin in the finite approach distance of ions to the surface (Stern layer concept), but can also relate to a potential drop within the carbon itself, called space charge effect, or quantum capacitance.[17,24] Note that dimensionless potentials, $\phi$, can be multiplied by the thermal voltage, $V_T$ (=$k_BT/e$) to arrive at voltages, $V$ or $E$, with units of Volts. At equilibrium, the electrode potential $E$ (in the carbon, relative to the solution phase outside the micropores) is given by

$$E/V_T = \Delta\phi_S + \Delta\phi_D, \tag{1}$$

where potentials $\Delta\phi_S$ and $\Delta\phi_D$ relate to the electronic and ionic charge density according to

$$\Delta\phi_S = \sigma_{elec} \cdot F/C_S \cdot V_T \quad , \quad \Delta\phi_D = -\text{arcsinh}(\sigma_{ionic}/2c_{salt}), \tag{2}$$

where $c_{salt}$ is the salt concentration outside the micropores. Note that all charge densities $\sigma_j$ are defined per unit micropore volume.

In the Donnan model, the total ion concentration in the micropores, $c_{ions,mi}$, is given by[7]

$$c_{ions,mi}^2 = \sigma_{ionic}^2 + 4 \cdot c_{salt}^2. \tag{3}$$

Eqs. (1)-(3) suffice to calculate the equilibrium ion adsorption in the micropores as function of electrode potential $E$, see Fig. 2. To describe a two-electrode CDI cell at equilibrium, we must evaluate Eqs. (1)-(3) for both electrodes, include that

$$V_{cell} = E_A - E_C \quad , \quad v_{mi,A} \cdot \sigma_{elec,A} + v_{mi,C} \cdot \sigma_{elec,C} = 0 \tag{4}$$

and analyze the CDI-cycle at two values of $V_{cell}$, namely the charging voltage, $V_{ch}$, and the discharge voltage, $V_{disch}$ (see Fig. 3). Comparing the values of $c_{ions,mi}$ in both electrodes at the end of the charging and discharge steps, multiplying by $v_{mi,j}$, summing over both electrodes, and multiplying by ½·$M_w$/$M_{elec}$ ($M_w$: molar mass of salt; $M_{elec}$: mass of both electrodes combined; in this manuscript $v_{mi}·M_w/M_{elec}$=7.4 mL/mol is used), results in the salt adsorption capacity, SAC, of the CDI cell, expressed in mg salt removal over the cycle per gram of both electrodes combined (when dry).[2]

A simple dynamic model can be set up that not only describes equilibrium salt adsorption but also the time-dependence of the salt concentration in the effluent water, see Fig. 4. To simplify the calculation, this model assumes that each electrode has properties invariant over its volume (no internal mass transfer limitations in the electrodes), while the ionic current is assumed proportional to an Ohmic potential drop, $\Delta\phi_\Omega$, between the two electrodes, i.e., across the spacer channel. More detailed theories for porous electrodes and spacers (including ion concentration variations) are given in refs. 1,2,4,22,25,26,27. The electronic current $I$ running through the external circuit from cathode to anode equals the ionic current running through the electrolyte in opposite direction. For electrodes without Faradaic reactions, a balance for the electronic charge in each electrode is given by

$$v_{mi,j} \cdot \frac{\partial \sigma_{elec,j}}{\partial t} = \pm I / F \tag{5}$$

where "+" is used for j=A and "-" for j=C. The ionic current relates to an Ohmic potential drop according to

$$I = \Delta V_\Omega / R_\Omega . \tag{6}$$

In the transport model, $\Delta V_\Omega$ is added to the right-hand side of Eq. (4a). In Eq. (6), $R_\Omega$ is an Ohmic transport resistance (dimension $\Omega$) which we assume is inversely proportional to the salt concentration, $c_{salt}$, according to $R_\Omega = \alpha / c_{salt}$.

Desalination of the water in the CDI cell is described by the salt balance

$$v_{tot} \cdot \frac{\partial c_{salt}}{\partial t} = \Phi \cdot \left( c_{salt,in} - c_{salt,out} \right) - \tfrac{1}{2} \cdot \left( J_{ions,A} + J_{ions,C} \right) \tag{7}$$

where $v_{tot}$ is the electrolyte volume in the cell excluding micropores, $\Phi$ the solution flowrate through the cell, $c_{salt,out}$ and $c_{salt,in}$ salt concentrations in the inflow and outflow of the cell, and $J_{ions,j}$ the ion flow rates from spacer to electrode micropores (in units of mol/s). In this "stirred tank" mass balance, Eq. (7), $c_{salt}$ is the salt concentration everywhere in the cell (except micropores), and thus also the salt concentration of the effluent ($c_{salt,out}=c_{salt}$). Finally, the ion flow rate into each electrode is given by

$$v_{mi,j} \cdot \frac{\partial c_{ions,mi,j}}{\partial t} = J_{ions,j} . \tag{8}$$

This set of differential algebraic equations suffices for a first-order model to simulate a CDI process that has fixed chemical charge in one or both electrodes.

In Fig. 2A, we present the results of the equilibrium model (Eqs. 1-3) applied to a single electrode, in order to study the effect of chemical charge on salt adsorption. We plot as function of the electrode potential, $E$, the electrode's electronic charge, $\sigma_{elec}$, and the total ion concentration in the micropores, $c_{ions,mi}$, for an unmodified electrode (solid curves), and for an electrode containing negative chemical charge (e.g., sulfonic groups, dashed curves). Such $c_{ions,mi}$ vs. $E$ plots have been used in previous works, but as schematics illustrating hypothesized electrode behavior.[11,12,14] Here, we directly derived such plots from an EDL model. Thus our single electrode $c_{ions,mi}$ vs. $E$ curves, for the first time in CDI, can be mathematically analyzed, quantitatively compared with data, and extended to non-equilibrium situations. In Fig. 2A, the minimum in the curve for $c_{ions,mi}$ corresponds to (the potential of) zero net ionic charge in the micropores, $\sigma_{ionic}=0$. At this potential $d\sigma_{elec}/dE$ is also at a minimum (but not $\sigma_{elec}$), because the ionic capacitance goes through a minimum when $\sigma_{ionic}=0$. For the unmodified electrode, this minimum corresponds to when $\sigma_{elec}=0$ and $E=0$, but for modified electrodes, it corresponds to where the electronic charge exactly compensates the chemical charge ($\sigma_{elec}+\sigma_{chem}=0$), a scenario schematically depicted in Fig. 1D. Note that for the unmodified electrode, the potential of zero charge (PZC) is the same when we consider ionic charge or electronic charge, in both cases the PZC is $E=0$. However, in the presence of chemical charge, the two PZCs are different, as $PZC_{elec}$ is located at -0.15 V, and $PZC_{ionic}$ at +0.6 V. Thus, when discussing the PZC,[12,14,24] our results demonstrate that it is important to clearly state whether we consider zero net *ionic* charge, $PZC_{ionic}$, or zero *electronic* charge, $PZC_{elec}$. The shift in the minimum in $c_{ions,mi}$ to higher potentials for electrodes with negative chemical charge (blue dashed curve) is in alignment with data for the differential capacitance of oxidized electrodes in Fig. 3 of ref. 14.

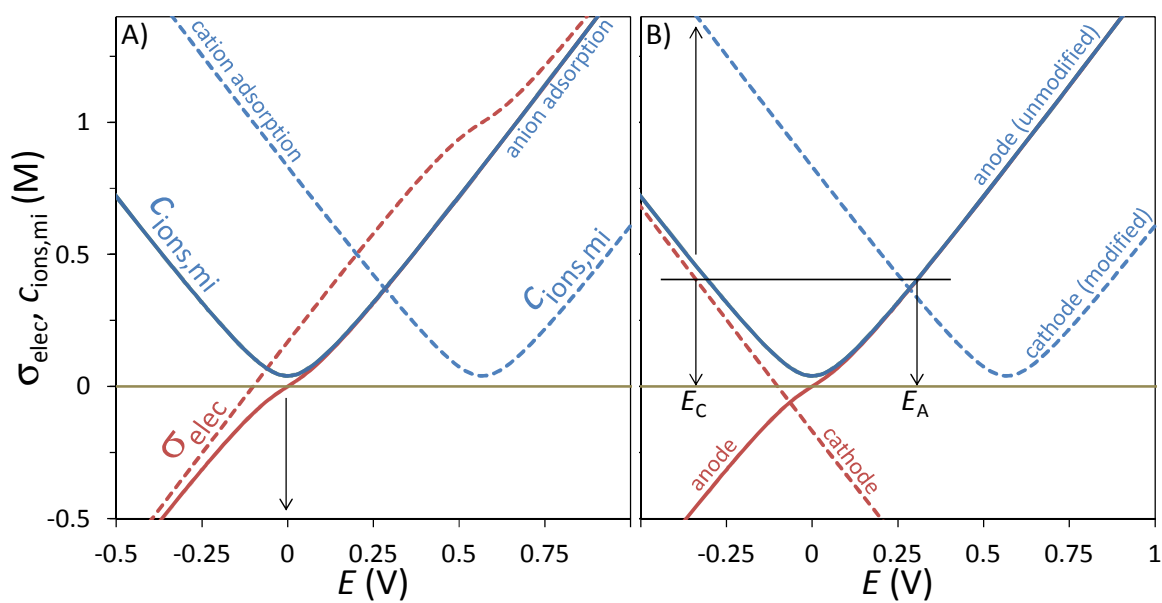

Fig. 2. A) Graphical analysis of single electrode behavior. Solid curves for unmodified electrode ($\sigma_{chem}=0$) and dashed curves for electrode containing negative chemical charge, $\sigma_{chem}=-1$ M ($c_{salt}=20$ mM, $C_S=170$ F/mL). In B) results for a two electrode cell, including an electrode modified as in A) (the cathode, dashed curves) and one unmodified electrode (the anode, solid curves).

In Fig. 2B, we provide an analysis of a two electrode CDI cell with equal micropore volumes ($v_{mi,A}=v_{mi,C}$, Eqs. 1-4) based on the two electrodes discussed in Fig. 2A (one modified, the other unmodified). Here, the modified electrode is the cathode in our two electrode cell (negatively polarized) and the unmodified electrode is the anode (positively polarized). Thus, the data in Fig. 2B is the same as that shown in Fig. 2A, except $\sigma_{elec}(E)$ for the modified electrode in Fig. 2B (red dashed line) was reflected in the x-axis compared to the same curve in Fig. 2A, as the value of $\sigma_{elec,C}$ must be equal and opposite to $\sigma_{elec,A}$ (Eq. 4b). This is done to facilitate graphical analysis, following the analysis shown in Fig. 8 of ref. 1. We analyze here one state, which is at the end of a hypothetical charging step where $\sigma_{elec,A}=-\sigma_{elec,C}\sim0.4$ M, and the resulting charging voltage is $V_{ch}=E_A-E_C\sim0.65$ V (see the horizontal black line in Fig. 2B). At both values of $E_j$ we can read off the ion concentration $c_{ions,mi,j}$ in the respective electrode. To analyze a CDI cycle, the same analysis must be performed for a second, discharge state (for example, at $\sigma_{elec,A}=\sigma_{elec,C}=0$) and to calculate SAC we must add up the values of $c_{ions,mi,A}$ and $c_{ions,mi,C}$ for the charged state, and subtracting from this $c_{ions,mi,A}+c_{ions,mi,C}$ at the discharge state, to arrive at the total ion adsorption in a cycle by the pair of electrodes, which can be converted to SAC in mg/g. This general analysis can be performed for a CDI cell with arbitrary values of the chemical charge in the two electrodes.

We now turn our attention to a second type of two electrode cell. Here, both electrodes again have equal micropore volumes ($v_{mi,A}=v_{mi,C}$), but now we set $\sigma_{chem,A}+\sigma_{chem,C}=0$, meaning the electrodes are modified such that they have equal magnitude but opposite sign of the chemical charges. This latter condition results in $c_{ions,mi}$ vs. $E$ curves for the anode and cathode that are mirror-images when reflected in a vertical axis about the point "$E=0$". Consequently, the anode potential $E_A$ can be multiplied by a factor of two to obtain the cell voltage, $V_{cell}=2 \cdot E_A$. For this cell, Eqs. (1)-(4) can be combined to the simple equation

$$V_{cell} = -2 \cdot V_T \cdot \left(\left(\sigma_{ionic,A} + \sigma_{chem,A}\right) \cdot F/C_S V_T + \text{arcsinh}\left(\sigma_{ionic,A}/2c_{salt}\right)\right), \tag{9}$$

which can be used along with Eq. (3) to construct $c_{ions,mi}$ vs. $V_{cell}$ curves, such as are shown in Fig. 3. As a comparison to our cell of interest with two modified electrodes, Fig. 3A shows the predicted curve for an unmodified pair of CDI electrodes ($\sigma_{chem,A}=\sigma_{chem,C}=0$; solid line). Here "0" denotes discharge at $V_{disch}=0$, and "1" charging at $V_{ch}=1.2$ V. The vertical distance between these two points is $\Delta c_{ions,mi}\sim0.85$ M, which results in a SAC of ~ 12.5 mg/g. In panel B we plot the same curve for our cell of interest, where the anode contains positive chemical charge, $\sigma_{chem,A}=+0.4$ M and the cathode negative charge of $\sigma_{chem,C}=-0.4$ M. In the anode the electronic charge is positive above a critical cell voltage of $V_{cell}^* = 2 \cdot V_T \cdot \text{arcsinh}\left(\sigma_{chem,A}/2c_{salt}\right) \approx 0.15$ V but switches to negative below $V_{cell}^*$; similarly, the cathode switches to carrying positive electronic charge below this threshold (see the black vertical line on Fig 3B). Comparing Fig. 3A and 3B, we note that cycling our modified cell between 0 and 1.2 V cell voltage (conditions "0" and "1") gives a somewhat increased value of $\Delta c_{ions,mi}$ and SAC (see inset of Fig. 3B for SAC values), similar to results in refs. 20 and 28. Now, if we instead cycle between charging at condition "1" and discharge at condition "2", which is at a negative discharge voltage, we observe an even higher $\Delta c_{ions,mi}$ and SAC. Because of the enhancement in SAC observed for this voltage range, we term this operating regime "enhanced CDI" or "e-CDI". If we now look at a third cycle, that between "0" and "3", we see that here the ion adsorption is lower at the end of the charging step (at "3") compared to discharge ("0"). This means that during cell charging salt transports from the electrode pores to the feedwater. The latter behavior of salt desorption during cell charging has been reported in the experimentally observed phenomenon of i-CDI.[10,11,13] As shown in the legend of Fig. 3B, the i-CDI cycle is also associated with a strong drop in SAC of the CDI cycle, to 1.2 mg/g compared to 14.6 mg/g for cycling between 0 and 1.2 V. Finally, the cycle between "0" and "4", while cycling between -1.2 and 0 V, demonstrates no inversion and a SAC of 3.4 mg/g, significantly lower than the case of "standard CDI" where we cycle between 0 and 1.2 V ("0" to "1"). Thus, we term this operational condition "diminished CDI". In summary, Fig. 3B demonstrates that with porous electrodes with chemical charge, the voltage range used in cell cycling is an exceptionally important parameter, as for a given cell this can dictate the operational regime of either standard CDI (here for $V_{cell}$=0 to 1.2 V), diminished CDI ($V_{cell}$=-1.2 to 0 V), i-CDI ($V_{cell}$=-0.8 to 0 V), and e-CDI ($V_{cell}$=-0.4 to 1.2 V). In the following paragraph, we will discuss how the performance regimes shown in Fig. 3B explain recent observations in experimental data.

To compare the results of Fig. 3B to recent experimental observations, we will first discuss the inversion phenomenon in i-CDI.[10,11,13] The latter phenomenon has been observed in CDI cells where at least one electrode is chemically modified and cell cycling is performed between $V_{ch}$~0.8 V and discharge at $V_{disch}$=0.[11,13] In Fig. 3B, we observe an i-CDI operational regime in our model cell for the identical voltage cycling between -0.8 and 0 V (note that in our model where the anode carries positive chemical charge, -0.8 V, and not 0.8 V, corresponds to inverted-CDI operation). The inversion phenomenon in our model is shown to occur also when cycling between conditions "0" and "2", i.e., at $V_{ch}$=-0.4 V, and this condition leads to a higher SAC magnitude than cycling between "0" and "3". A second operating regime seen in our model is that of "diminished CDI", and recent experimental results have also reported a decrease in SAC for the case when during long-term cycling negative chemical charge develops in the positively polarized electrode of a CDI cell (and, we assume, opposite charge in the other electrode), corresponding to cycling between "0" and "4" in Fig. 3B.[10,11] It is also reported that reversing the polarity of such electrodes (cathodic polarization) leads to restoration of SAC, which in our model would correspond to cycling between "0" and "1" (standard CDI),[10,12] and thus these literature observations are in agreement with Fig. 3B. In summary, Fig. 3B captures the essential features reported in two operating regimes observed in recent experimental CDI cells (i-CDI and diminished CDI), as well as the response to cathodic polarization.

The third operating regime demonstrated by our model in Fig. 3B, that of e-CDI, has not yet been demonstrated experimentally or theoretically predicted, to our knowledge. The operational regime of e-CDI would significantly advance the field of ion electrosorption, as capacitive electrodes have until now been able to attain roughly 14 mg/g SAC, and our model results show that significant enhancements to this value may be possible.[2] Using a discharge voltage of $V_{disch}$~-0.4 V in our model, SAC significantly increased, to a maximum of ~18.2 mg/g, which compared to the maximum for unmodified electrodes (12.5 m/g, see Fig 3A) is an increase of ~50%. This improvement is obtained at a still rather small chemical charge of ±0.4 M. Commercial ion-exchange membranes have fixed charge densities in excess of 5 M per unit aqueous pore volume in the membrane[23] and thus higher values than 0.4 M may be experimentally feasible and will further boost the capacity of carbon electrodes by leveraging the e-CDI operational regime. The optimum value of $V_{disch}$ is found when electronic and fixed chemical charge exactly charge-compensate and thus (for $\sigma_{chem,A}+\sigma_{chem,C}$=0) $V_{disch}$ depends on the chemical charge density according to $V_{disch,opt} = -2 \cdot \sigma_{chem,A} \cdot F / C_S$. If the capacitance $C_S$ is known, this equation predicts the optimum chemical charge to have operation of e-CDI utilizing the full span of the water stability window, e.g., for $C_S$=170 F/mL, to have $V_{disch,opt}$=-1.23 V we predict that $\sigma_{chem,A}$ must be ~1.1 M. Vice-versa, this equation can also be used to derive $\sigma_{chem,A}$ from knowledge of $C_S$ and of the measured optimum value of $V_{disch}$.

The above analysis considers symmetrically modified electrodes ($\sigma_{chem,A}+\sigma_{chem,C}=0$). However, for the asymmetric case, e.g., if the anode has fixed positive chemical charge, and the cathode is unmodified, the enhancement possible with e-CDI is much reduced; in effect, the least-charged electrode is limiting the process. This is shown in Fig. 3A by the dashed curve ($\sigma_{chem,A}=0.4$ M, $\sigma_{chem,C}=0$) which shows that any choice of cycle voltages, $V_{ch}$ and $V_{disch}$, results in the asymmetrically modified system having a lower value of $\Delta c_{ions,mi}$ than the unmodified cell.

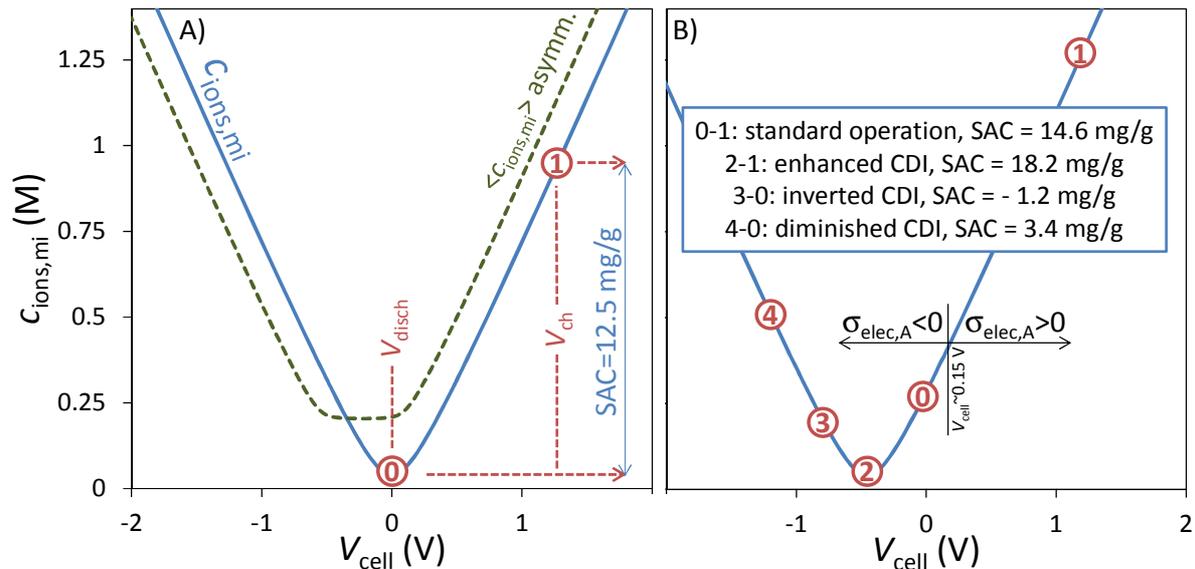

Fig. 3. Graphical analysis of two-electrode behavior (symmetric case where $\sigma_{chem,A}=-\sigma_{chem,C}=0.4$ M, $c_{salt}=20$ mM). A) Solid curve: classical CDI with unmodified electrodes, where $c_{ions,mi}$ is symmetric around $V_{cell}=0$. Charging at condition "1" ($V_{ch}=1.2$ V) and discharging at "0" ($V_{disch}=0$) results in a salt adsorption capacity SAC as indicated. Reversing the value of $V_{ch}$ gives the same SAC. Green dashed line: asymmetrically modified electrodes ($\sigma_{chem,A}=0.4$ M; $\sigma_{chem,C}=0$). B) With chemically modified electrodes, charging at +1.2 V (discharge at $V_{disch}=0$) gives a slightly higher SAC as in A) but charging at -1.2 V (condition "4") results in a significantly reduced SAC. i-CDI is observed when cycling between discharge at "0" and charging at either "2" or "3".

While Fig. 3 and Eq. (9) demonstrate the equilibrium salt sorption of unmodified and modified CDI cells, in Fig. 4, we show the desalination dynamics for these same two cells (Eqs. 5-8). In particular, dynamic calculations of the effluent salinity vs. time can be compared with experimental data reported for cells using electrodes with chemical charge, such as those demonstrating i-CDI. Calculation results are based on $C_S=170$ F/mL, $v_{tot}/v_{mi}=10$, residence time $v_{tot}/\Phi=16$ s, and a transport resistance $\alpha \cdot v_{mi}=1$ μΩ·mol, and, as can be seen in Fig. 4, using these parameters we observe unique dynamical features which are similar to reported behavior in experimental i-CDI cells.[10-13] In Fig. 4, panel A shows model results obtained with unmodified electrodes ($\sigma_{chem,A}=\sigma_{chem,C}=0$), and we can observe that the results show the classical behavior of a CDI cell with a decreasing value of $c_{salt,out}$ during charging (which is the period 0-100 s), and subsequent salt release during discharge (period 100-200 s). In panels B-D, we show calculation results with modified electrodes ($\sigma_{chem,A}=-\sigma_{chem,C}=0.4$ M). Panel B shows a cell in the diminished CDI operational regime cycling between -1.2/0 with a peculiar salination peak observed at the start of the charging step. Interestingly, this peak has been observed to develop during prolonged cycling of a CDI cell and called a "repulsion peak",[10] or "inversion peak".[12] At -0.8/0 we show theoretically i-CDI behavior with salt release during charging (first period) and desalination during discharge; at -0.4/0 this inversion is even more prominent. Thus, our simple dynamical model reproduces the essential and critical features observed in refs. 10-13 when using chemically charged electrode(s) for CDI, including inversion behavior in certain charge/discharge voltage ranges, an inversion peak, as well as reduced SAC.

In conclusion, our EDL model extended with chemical charge describes the most pertinent experimental observations related to inverted CDI. This agreement strongly supports the basic assumption that chemical charge residing in the micropores of i-CDI electrodes is responsible for the inversion phenomenon. Beyond i-CDI, we show that chemical charge in porous CDI electrodes can also lead to a multitude of operational regimes some of which can be detrimental for CDI operation, and others which can be used to enhance desalination significantly (e-CDI). Future experimental and theoretical work should establish the concentration of chemical charge as function of synthesis method, and further investigate the validity of the Donnan model for modified electrodes, possible transport limitations in these electrodes, and effects of chemical charge on the window of operation safe from water electrolysis.

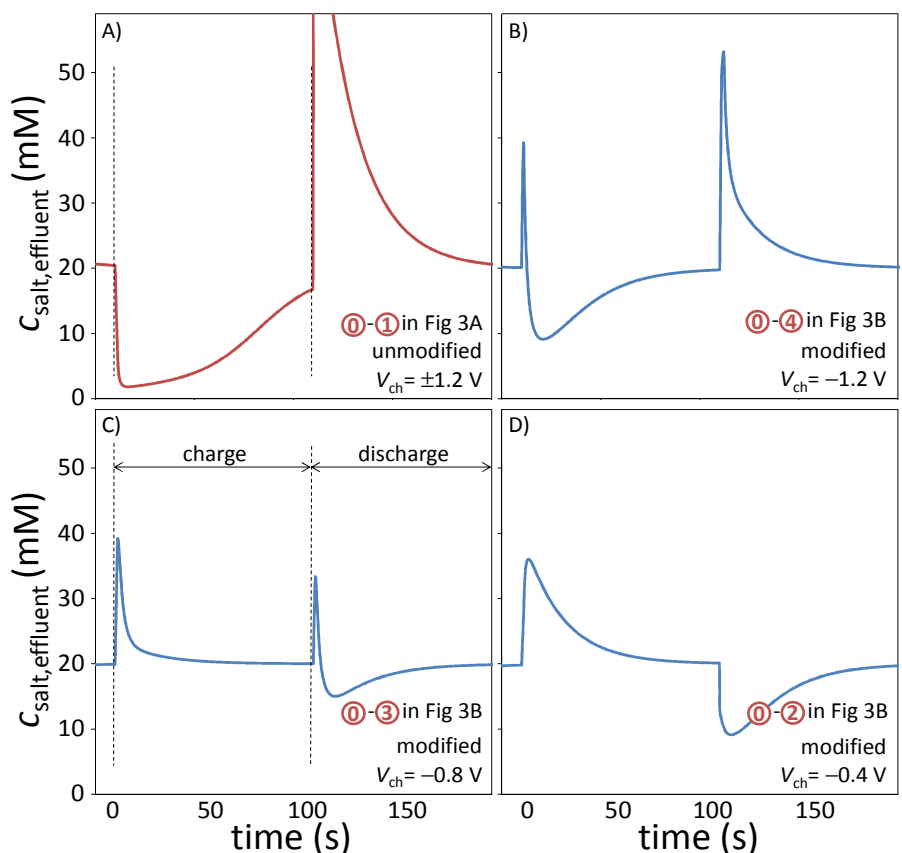

Fig. 4. Effluent salt concentration in CDI, for a charging cycle (from 0 to 100 s) and discharge cycle (100 to 200 s, always $V_{disch}=0$, $c_{salt}=20$ mM). A) Classical CDI with unmodified electrodes at $V_{ch}=\pm1.2$ V. B-D) CDI with electrodes containing chemical charge ($\sigma_{chem,A}=-\sigma_{chem,C}=0.4$ M) for decreasing values of $V_{ch}$. In panels C) and D), the inversion phenomenon is observed, while in B) the repulsion peak is found at the start of a charging step (around 0 s).

**Acknowledgments**

This work was performed in the cooperation framework of Wetsus, European Centre of Excellence for Sustainable Water Technology (www.wetsus.eu). Wetsus is co-funded by the Dutch Ministry of Economic Affairs and Ministry of Infrastructure and Environment, the Province of Fryslân, and the Northern Netherlands Provinces. We thank Xin Gao, James Landon and Ayokunle Omosebi (University of Kentucky, USA) for useful suggestions during preparation of this manuscript.

**References**

1. A.M. Johnson and J. Newman, *J. Electrochem. Soc.* **118** (1971) 510.
2. M.E. Suss, S. Porada, X. Sun, P.M. Biesheuvel, J. Yoon, and V. Presser, *Energy & Env. Sci.* (2015), DOI:10.1039/c5ee00519a.
3. R.A. Rica, R. Ziano, D. Salerno, F. Mantegazza, D. Brogioli, *Phys. Rev. Lett.* **109** (2012) 156103.
4. M. Mirzadeh, F. Gibou, and T.M. Squires, *Phys. Rev. Lett.* **113** (2014) 097701.
5. M. Müller and B. Kastening, *J. Electroanal. Chem.* **374** (1994) 149.
6. B. Kastening and M. Heins, *Electrochimica Acta* **50** (2005) 2487.
7. P.M. Biesheuvel, S. Porada, M. Levi and M.Z. Bazant, *J. Solid State Electrochem.* **18** 1365 (2014).
8. B. Giera, N.J. Henson, E.M. Kober, M.S. Shell, and T.M. Squires, *Langmuir* **31** (2015) 3553.
9. C. Prehal, D. Weingarth, E. Perre, R.T. Lechner, H. Amenitsch, O. Paris, and V. Presser, *Energy & Environm. Sci.* **8** (2015) 1725.
10. Y. Bouhadana, E. Avraham, M. Noked, M. Ben-Tzion, A. Soffer, and D. Aurbach, *J. Phys. Chem. C* **115** (2011) 16567.
11. I. Cohen, E. Avraham, Y. Bouhadana, A. Soffer, and D. Aurbach, *Electrochim. Acta* **106** (2013) 91.
12. X. Gao, A. Omosebi, J. Landon, and K. Liu, *J. Electrochem. Soc.* **161** (2014) E159.
13. X. Gao, A. Omosebi, J. Landon, and K. Liu, *Energy & Environm. Sci.* **8** (2015) 897.
14. A. Omosebi, X. Gao, J. Rentschler, J. Landon, and K. Liu, *J. Colloid Interface Sci.* **446** (2015) 345.
15. H. Gerischer, *Adv. Electrochem. Electrochem. Eng.* **1**, P. Delahay, Interscience (1961) 139.
16. R. Memming and G. Schwandt, *Angew. Chem. Intl. Ed.* **6** (1967) 851.
17. H. Gerischer, R. McIntyre, D. Scherson, and W. Storck, *J. Phys. Chem.* **91** (1987) 1930.
18. Z. Jiang and D. Stein, *Phys. Rev. E* **83** (2011) 031203.
19. M.E. Fleharty, F. van Swol, and D.N. Petsev, *Phys. Rev. Lett.* **113** (2014) 158302.
20. J. Yang, L. Zou, and N.R. Choudhury, *Electrochim. Acta* **91** (2013) 11.
21. M. Marino *et al.*, *J. Colloid Interface Sci.* **436** (2014) 146.
22. T. Kim, J.E. Dykstra, S. Porada, A. van der Wal, J. Yoon, and P.M. Biesheuvel, *J. Colloid Interface Sci.* **446** (2015) 317.
23. A.H. Galama, J.W. Post, M.A. Cohen Stuart, P.M. Biesheuvel, *J. Membrane Sci.* **442** (2013) 131.
24. L.-H. Shao *et al.*, *PCCP* **12** (2010) 7580.
25. M.E. Suss, P.M. Biesheuvel, Th.F. Baumann, M. Stadermann, and J.G. Santiago, *Environ. Sci. Technol.* **48** (2014) 2008.
26. R.A. Rica, R. Ziano, D. Salerno, F. Mantegazza, M.Z. Bazant, and D. Brogioli, *Electrochim. Acta* **92** (2013) 304.
27. J. Gabitto and C. Tsouris, *Transp. Porous Med.* (2015) DOI:10.1007/s11242-015-0502-0.
28. D. Andelman, *J. Mater. Sci. Chem. Eng.* **2** (2014) 16.